\documentclass[a4paper,11pt]{article}
\usepackage{pos}
\usepackage{subcaption}
\usepackage{setspace}
\usepackage{lineno}

\title{Studies of systematic uncertainty effects on IceCube's real-time angular uncertainty}
 \ShortTitle{Studies of systematic uncertainty effects}

\author{The IceCube Collaboration \\{\normalsize \normalfont(a complete list of authors can be found at the end of the proceedings) \vspace{2mm}}}

\emailAdd{cristina.lagunas@icecube.wisc.edu}
\emailAdd{yosuke.ashida@icecube.wisc.edu}
\emailAdd{ankur.sharma@icecube.wisc.edu}
\emailAdd{hamish.thomas@icecube.wisc.edu}

\abstract{

Sources of astrophysical neutrinos can potentially be discovered through the detection of neutrinos in coincidence with electromagnetic or gravitational waves. Real-time alerts generated by IceCube play an important role in this search, acting as triggers for follow-up observations with instruments sensitive to other wavelengths. Once a high-energy event is detected by the IceCube real-time program, a complex and time consuming direction reconstruction method is run in order to calculate an accurate localisation. To investigate the effect of systematic uncertainties on the uncertainty estimate of the location, we simulate a set of high-energy events with a wide range of directions for different ice model realisations, the dominant systematic error in our localization uncertainty. This makes use of a novel simulation tool, which allows the treatment of systematic uncertainties with multiple continuously varied nuisance parameters. These events will be reconstructed using various reconstruction methods. This study will enable us to include systematic uncertainties in a robust manner in the real-time direction and error estimates. \\

\vspace{4mm}
{\bfseries Corresponding authors:}
Cristina Lagunas Gualda$^{1*}$, Yosuke Ashida$^{2}$, Ankur Sharma$^{3}$, Hamish Thomas$^{4}$\\
{$^{1}$ \itshape DESY}\\
{$^{2}$ \itshape University of Wisconsin-Madison}\\
{$^{3}$ \itshape Uppsala University}\\
{$^{4}$ \itshape University of Canterbury}\\[4mm]
$^*$ Presenter

\FullConference{37$^{\rm{th}}$ International Cosmic Ray Conference (ICRC 2021)\\
		July 12th -- 23rd, 2021\\
		Online -- Berlin, Germany}
}

\begin{document}
\maketitle

\section{Introduction}

IceCube is a cubic-kilometer neutrino detector installed in ice at the geographic South Pole~\cite{Aartsen:2016nxy} between the depths of 1450 m and 2450 m. Reconstruction of the direction, energy and flavor of neutrinos relies on the optical detection of Cherenkov radiation emitted by secondary charged particles produced in the interactions of neutrinos with the surrounding ice or the nearby bedrock. The main body of the detector is a 3-D array of 5160 Digital Optical Modules (DOMs) that contain the photomultipliers that detect the Cherenkov photons. 

The real-time analysis framework~\cite{Aartsen:2016lmt} was implemented in 2017 with the aim of identifying the sources of astrophysical neutrinos by detecting events in coincidence with electromagnetic or gravitational waves. Part of this program comprises real-time alerts generated by IceCube: identifying very-high-energy neutrinos with a high probability of being of astrophysical origin. This framework was crucial in the identification of a high-event neutrino event, IC170922A, in spatial and temporal coincidence with a gamma-ray flux from the blazar TXS 0506+056 with $3\sigma$ significance~\cite{IceCube:2018dnn}. 

The pipeline works as follows: once a neutrino is detected, and if it passes the threshold to become an IceCube alert~\cite{Blaufuss:2019fgv}, it is reconstructed at the South Pole using a computationally limited reconstruction method included in the online processing and filtering system~\cite{Aartsen:2016nxy}. A first GCN Notice is sent out with this information, including the statistical angular uncertainty on the localization. Immediately after, a more robust and sophisticated reconstruction method is applied to the data. A brute force scan of all possible directions is performed pixelwise in order to produce a likelihood landscape, returning the best-fit position as the pixel with the minimum likelihood, $\mathcal{L}_0$. This method allows us to obtain error contours that account for systematic uncertainties. The resulting direction (including statistical and systematic errors) from the scan is sent out in a second GCN Notice and a GCN Circular. 

Since Wilks' theorem does not apply, Monte Carlo simulations are needed to calibrate the error contours derived from the likelihood landscape in order to account for statistical and systematic uncertainties. A neutrino event is re-simulated and reconstructed to calculate the likelihood values that correspond to a given containment. These correction values are applied to every IceCube alert to scale the contours, regardless of topology of the track. The aim of this study is to verify this scaling and create an array of correction values to choose from for each neutrino event. Section~\ref{sec:reco} introduces the reconstruction method considered in this analysis. In section~\ref{sec:previous} the calculation of the currently used correction values is explained. In section~\ref{sec:simulation} the simulation of new muons that are used to study the validity of those values is discussed. Section~\ref{sec:status} contains the status of this work. Lastly, we present the future plans in section~\ref{sec:outlook}.

\section{Reconstruction}\label{sec:reco}

\textit{Millipede} is a reconstruction algorithm designed to infer muon energy losses from the deposited light in the detector~\cite{Aartsen:2013vja}. This method assumes that the light emission from a high-energy muon can be described by a series of cascades whose energies ($E_i$) can be estimated. The number of detected photons~$k$ is expected to follow a Poisson distribution with the mean $\lambda$ given by

\begin{equation}
    \lambda  = \sum_{\mathrm{sources}\ i} E_i \Lambda_i + \rho, 
\end{equation}

\noindent 
where $\Lambda_i$ is the expected light yield in a particular DOM and time bin from cascade $i$, and $\rho$ is the expected number of noise photons. Assuming a Poisson distribution, the resulting likelihood summed over time bins $j$ that has to be minimized to find the best-fit $\overrightarrow{E}$ is

\begin{equation}
    -\sum_j \log \mathcal{L}  =- \sum_j k_j \log \left(\overrightarrow{E} \cdot \overrightarrow{\Lambda}_j + \rho_j\right) + \sum_j \log \left(\overrightarrow{E} \cdot \overrightarrow{\Lambda}_j + \rho_j\right) + \sum_j \left( \log k_j! \right). 
\end{equation}

This method can also be used to reconstruct the direction of high-energy muons. Firstly, the best-fit energy losses and the likelihood are calculated for a fixed direction. Secondly, the direction and the reference location are varied and those parameters are re-calculated. These two steps can be repeated until the global minimum likelihood is found.

In practice, this is done on a pixel grid using \textit{Hierarchical Equal Area isoLatitude Pixelation of a sphere} (\textit{healpix})~\cite{Gorski_2005}. In a typical scan, the full sky is divided into pixels of equal area, with iteratively finer binning near the minimum. In this study, the scan is focused on a small area of the likelihood landscape around the known true direction with the finest binning used in the real-time program.

\section{Current correction values}\label{sec:previous}

Monte Carlo re-simulations were done for the first time for the high-energy track IC160427A, for which a possible counterpart supernova was found by Pan-STARRS~\cite{Kankare:2019bzi}. Re-simulations consist of simulating muons that are ``similar'' to one specific event, varying the parameters of the ice model used to propagate particles. In this first analysis, similarity was defined as difference in the true direction of ±2 degrees, distance to the original track below 50 m and ±20\% of the charge deposited in the detector by the photons. This choice ensured that every simulated event closely resembled the original neutrino. The systematic uncertainties were included by sampling parameters from a predefined ice model and varying them with a Gaussian distribution during the photon propagation.

\begin{figure}[h]
    \centering
    \includegraphics[trim={0 20 0 0},clip, width=0.6\textwidth]{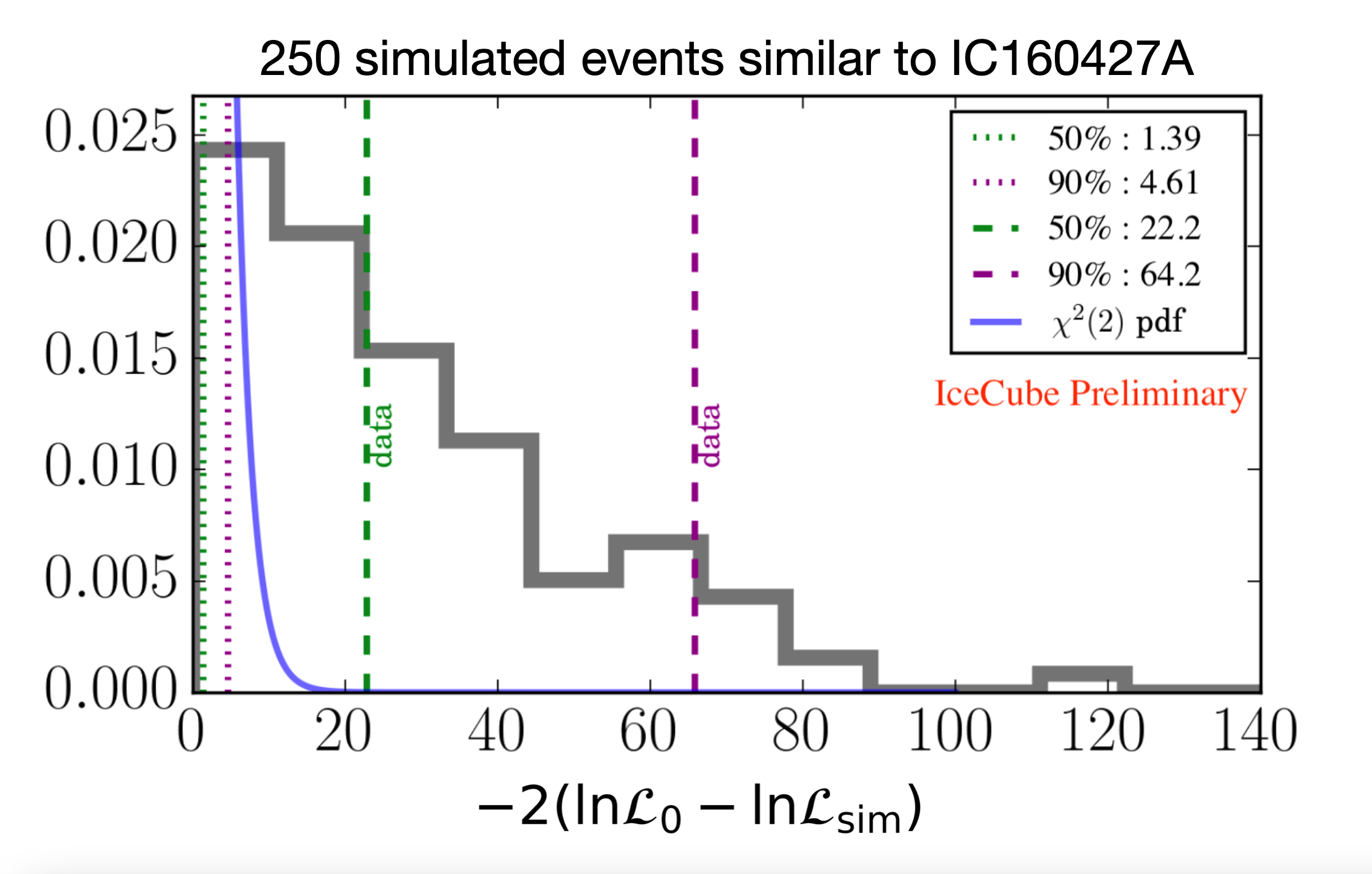}
    \caption{Distribution of the difference between the likelihood of the \textit{Millipede} scan best fit ($\mathcal{L}_{0}$) and the likelihood of the true direction ($\mathcal{L}_{\mathrm{sim}}$) for 250 re-simulations of the IceCube's neutrino alert IC160427A. The 50\% (90\%) containment is marked with a green (magenta) dashed line and the value is included in the legend. The $\chi^2$ distribution that the data would follow according to Wilks' theorem is also shown (blue solid line), along with its corresponding 50\% and 90\% containment (dotted lines).}
    \label{fig:160427aResim}
\end{figure}

Subsequently, the muons were reconstructed using the \textit{Millipede} reconstruction method. The ratio between the likelihood of the best fit ($\mathcal{L}_{0}$) and the Monte Carlo truth (simulated direction) ($\mathcal{L}_{\mathrm{sim}}$) can be used to construct a histogram from which one can extract the correction values. Figure \ref{fig:160427aResim} shows this distribution for 250 events similar to IC160427A. The correction values are the~$-2\left(\log \mathcal{L}_{0} - \log \mathcal{L}_{\mathrm{sim}}\right)$ values that correspond to 50\%/90\% containment. In this case, they are 22.2 and 64.2, respectively. These are currently used to convert the likelihood landscape of the \textit{Millipede} scans to error contours for every alert by searching for the pixels that satisfy~$-2\left(\log \mathcal{L}_{0} - \log \mathcal{L}_{\mathrm{pixel}}\right) = 22.2 (64.2)$.

With the analysis presented here, we investigate if this single set of correction values is enough to represent all the possible neutrino events that are detected in IceCube, i.e. if the impact of the ice systematic uncertainties on the resolution of the reconstruction is the same throughout the detector.

\section{Simulation}\label{sec:simulation}

\begin{figure}[h]
    \centering
    \includegraphics[width=0.4\textwidth]{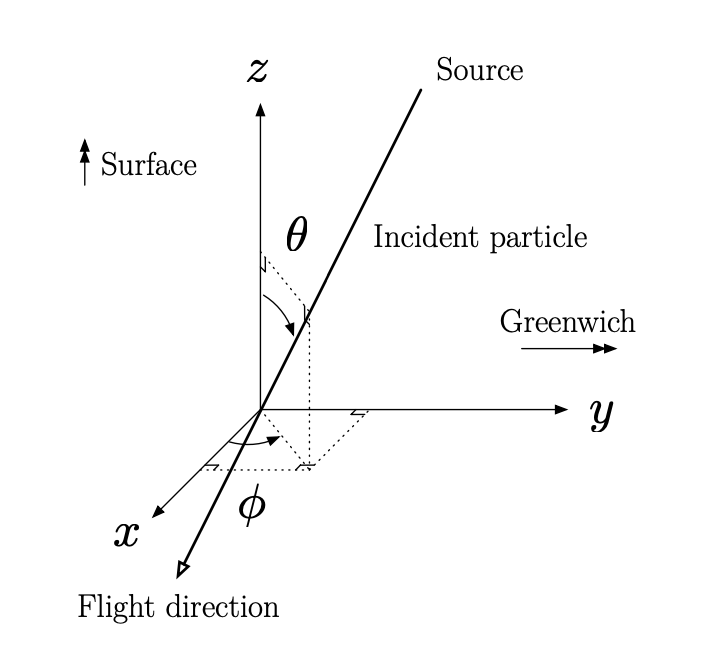}
    \caption{The IceCube coordinate system, centered 1948 m below the surface of the glacier.}
    \label{fig:coordsys}
\end{figure}

\begin{figure}
     \centering
     \begin{subfigure}[b]{0.3\textwidth}
         \centering
         \includegraphics[width=\textwidth]{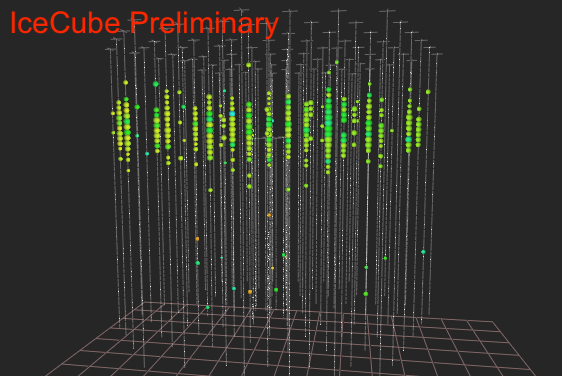}
         \caption{\textit{Horizontal Shallow Smooth}}
         \label{fig:ev1}
     \end{subfigure}
     \hfill 
     \begin{subfigure}[b]{0.3\textwidth}
         \centering
         \includegraphics[width=\textwidth]{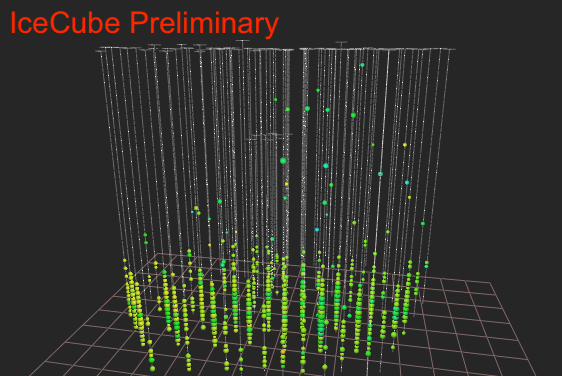}
         \caption{\textit{Horizontal Deep Smooth}}
         \label{fig:ev2}
     \end{subfigure}
     \hfill
     \begin{subfigure}[b]{0.3\textwidth}
         \centering
         \includegraphics[width=\textwidth]{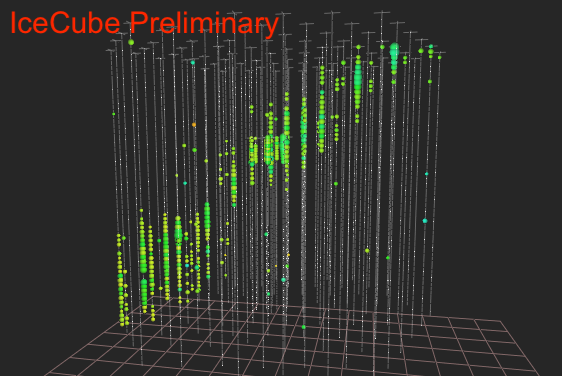}
         \caption{\textit{Upgoing Smooth}}
         \label{fig:ev3}
     \end{subfigure}
     \hfill
     \begin{subfigure}[b]{0.3\textwidth}
         \centering
         \includegraphics[width=\textwidth]{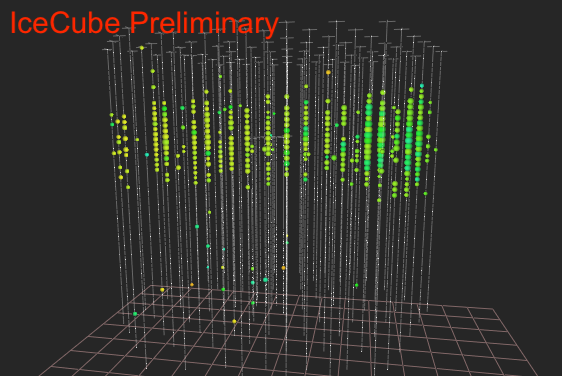}
         \caption{\textit{Horizontal Shallow Stochastic}}
         \label{fig:ev4}
     \end{subfigure}
     \hfill
     \begin{subfigure}[b]{0.3\textwidth}
         \centering
         \includegraphics[width=\textwidth]{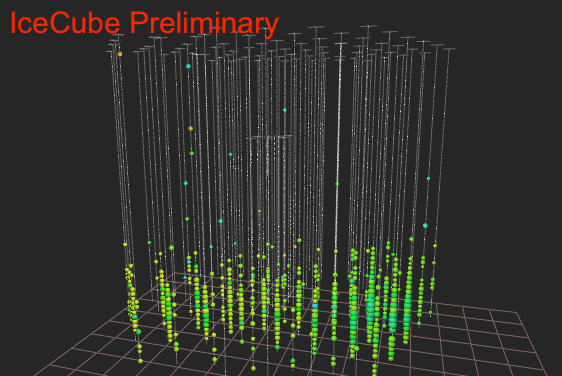}
         \caption{\textit{Horizontal Deep Stochastic}}
         \label{fig:ev5}
     \end{subfigure}
     \hfill
     \begin{subfigure}[b]{0.3\textwidth}
         \centering
         \includegraphics[width=\textwidth]{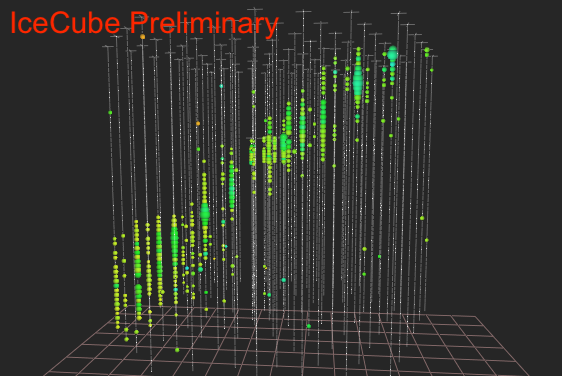}
         \caption{\textit{Upgoing Stochastic}}
         \label{fig:ev6}
     \end{subfigure}
        \caption{Event views of the 6 types of re-simulations.}
        \label{fig:eventviews}
\end{figure}

Since our first goal was to determine the validity of the correction values calculated with IC160427A, simple categories of muons that represent the majority of alerts are defined. All of these neutrino-candidate events are categorized as through-going tracks with energies close to the median energy of IceCube's alerts (150~TeV). Two main classes exist: \textit{Horizontal} ($\theta = 90$~degrees, see Figure \ref{fig:coordsys}) and \textit{Upgoing} ($\theta = 130$~degrees) muons. Due to the filters in the event selection, most neutrino alerts have zenith angles between 80 and 140 degrees. The impact of stochasticity is studied by choosing a \textit{Smooth} muon (with continuous energy losses along the track) and a \textit{Stochastic} muon (at least one big stochastic energy loss in the track) for each main category. The distinction is based on the calculation of the highest energy loss divided by the median energy loss along the track. For the \textit{Horizontal} class two depths in the detector were selected to account for differences in ice properties, one in the upper half at $z = 200$~m (\textit{Shallow}) and the other in the lower half at $z = -400$~m (\textit{Deep}). That leaves us with a total of 6 types of muons (Figure~\ref{fig:eventviews} for event views).

100 muons for each of the categories defined above are simulated. The photon propagation step is repeated utilising a novel simulation tool called \textit{SnowStorm}~\cite{Aartsen:2019jcj} resulting in varying ice systematics between muons. This differs from the previous analysis in two ways: firstly, \textit{SnowStorm} has a more robust treatment of systematic uncertainties than what was done previously. Secondly, the direction and deposited charge are fixed, which allows us to better understand the impact of the ice model parameters on the reconstruction angular uncertainty.

\section{Current status}\label{sec:status}

\begin{figure}[h]
    \centering
    \includegraphics[width=0.6\textwidth]{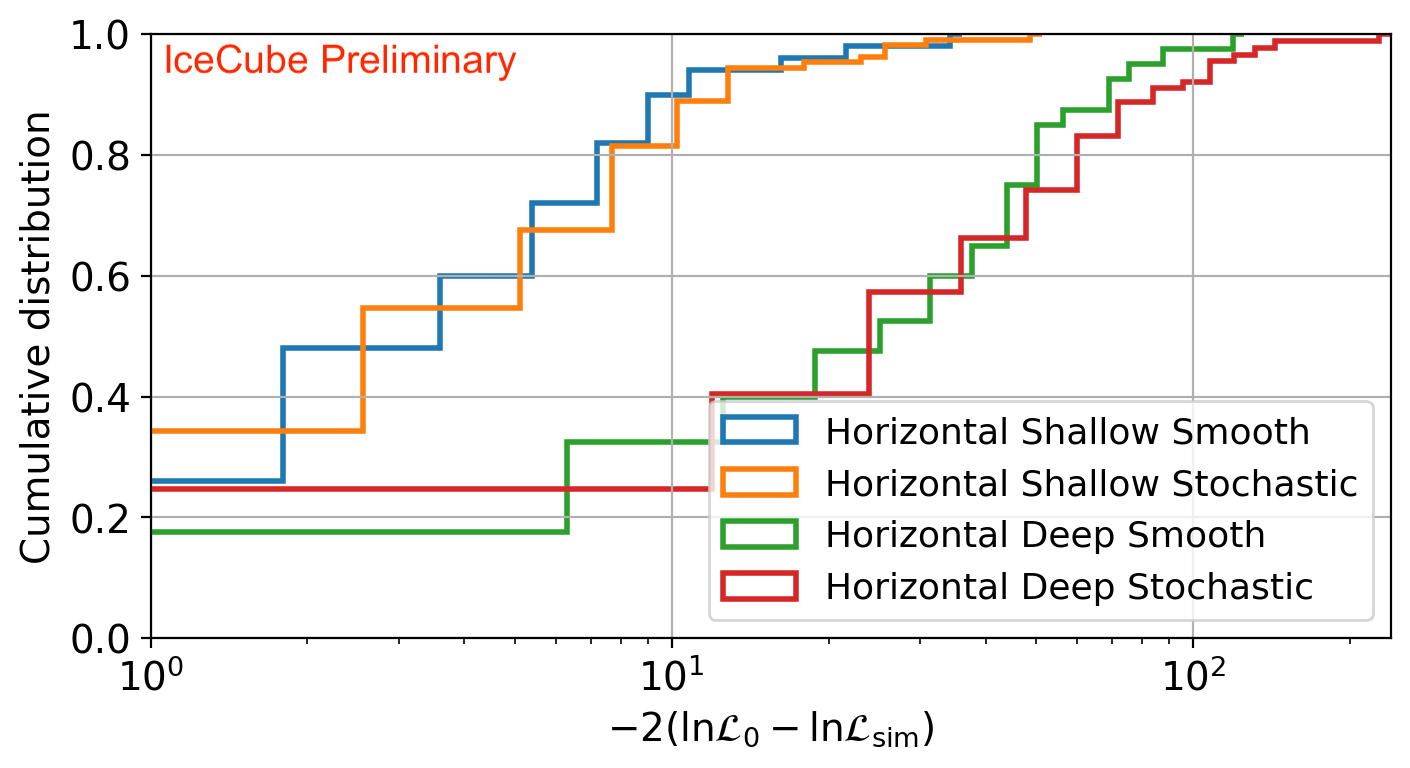}
    \caption{Cumulative distribution of the difference between the likelihood of the \textit{Millipede} scan best fit ($\mathcal{L}_{0}$) and the likelihood of the true direction ($\mathcal{L}_{\mathrm{sim}}$) for the \textit{Horizontal} muons. The two \textit{Smooth} and \textit{Stochastic} classes are visually compatible, which is confirmed with a Kolmogorov-Smirnov test.}
    \label{fig:cumulativeHorizontal}
\end{figure}

The first result obtained is shown in Figure~\ref{fig:cumulativeHorizontal}, where the cumulative distribution of the difference in likelihood between the best-fit positions and the simulated directions of the \textit{Horizontal} muons is plotted. One can visually see that stochasticity does not play an important role in the calibration of the reconstruction for these categories. A Kolmogorov-Smirnov test was conducted to check the compatibility of the distributions resulting in a p-value of 0.67 (0.65) for the \textit{Smooth} (\textit{Stochastic}) muons. Therefore, the categories are merged into \textit{Horizontal Shallow} and \textit{Horizontal Deep}. 

\begin{figure}[h]
    \centering
    \includegraphics[width=0.6\textwidth]{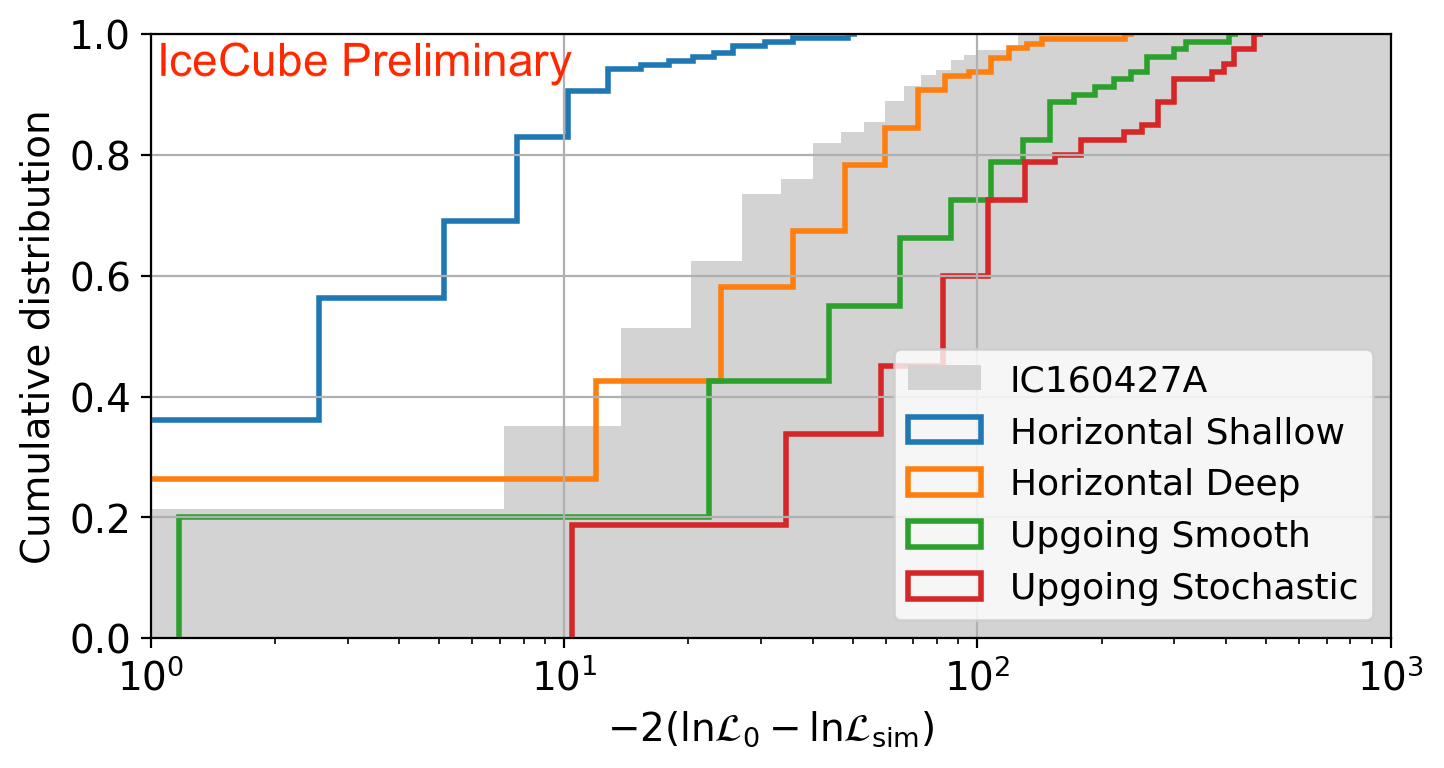}
    \caption{Cumulative distribution of the difference between the likelihood of the \textit{Millipede} scan best fit ($\mathcal{L}_{0}$) and the likelihood of the true direction ($\mathcal{L}_{\mathrm{sim}}$) for four selected categories of simulated muons including systematic uncertainties. For comparison, the cumulative distribution for IC160427A is included in gray.}
    \label{fig:cumulativeGM}
\end{figure}

In Figure \ref{fig:cumulativeGM}, the cumulative distribution of the difference in likelihood between the best-fit positions and the simulated directions for the four selected categories is shown. Table \ref{tab:status} shows the correction values and the number of reconstructed events. Despite a difference in the number of re-simulations, one can clearly see that each class has a unique set of values; the impact of systematic uncertainties depends on the position and shape of the track in the detector. This strongly implies that it is not ideal to calculate the error contours using the re-simulations of IC160427A for every neutrino alert.

\begin{table}[h]
\centering
\begin{tabular}{l c c c}
\hline \hline
Category               & 50\% containment & 90\% containment & Number of events \\ \hline 
\textit{Horizontal Shallow} & 4.5  & 12.5  & 158              \\
\textit{Horizontal Deep}    & 31.9 & 83.2  & 129              \\ 
\textit{Upgoing Smooth}     & 51.8 & 193.8 & 80               \\
\textit{Upgoing Stochastic} & 88.9 & 301.7 & 80               \\ \hline 
IC160427A                   & 22.2 & 64.2  & 250              \\ \hline \hline
\end{tabular}
\caption{The correction values for the selected categories, along with the number of reconstructed events at the time of writing this proceeding. Some of the simulated events are not included as they have not yet been reconstructed at the time of writing. For comparison, the values from the IC160427A re-simulations are also shown.}
\label{tab:status}
\end{table}

\begin{figure}[h]
\centering
\begin{subfigure}{.5\textwidth}
  \centering
  \includegraphics[width=\linewidth]{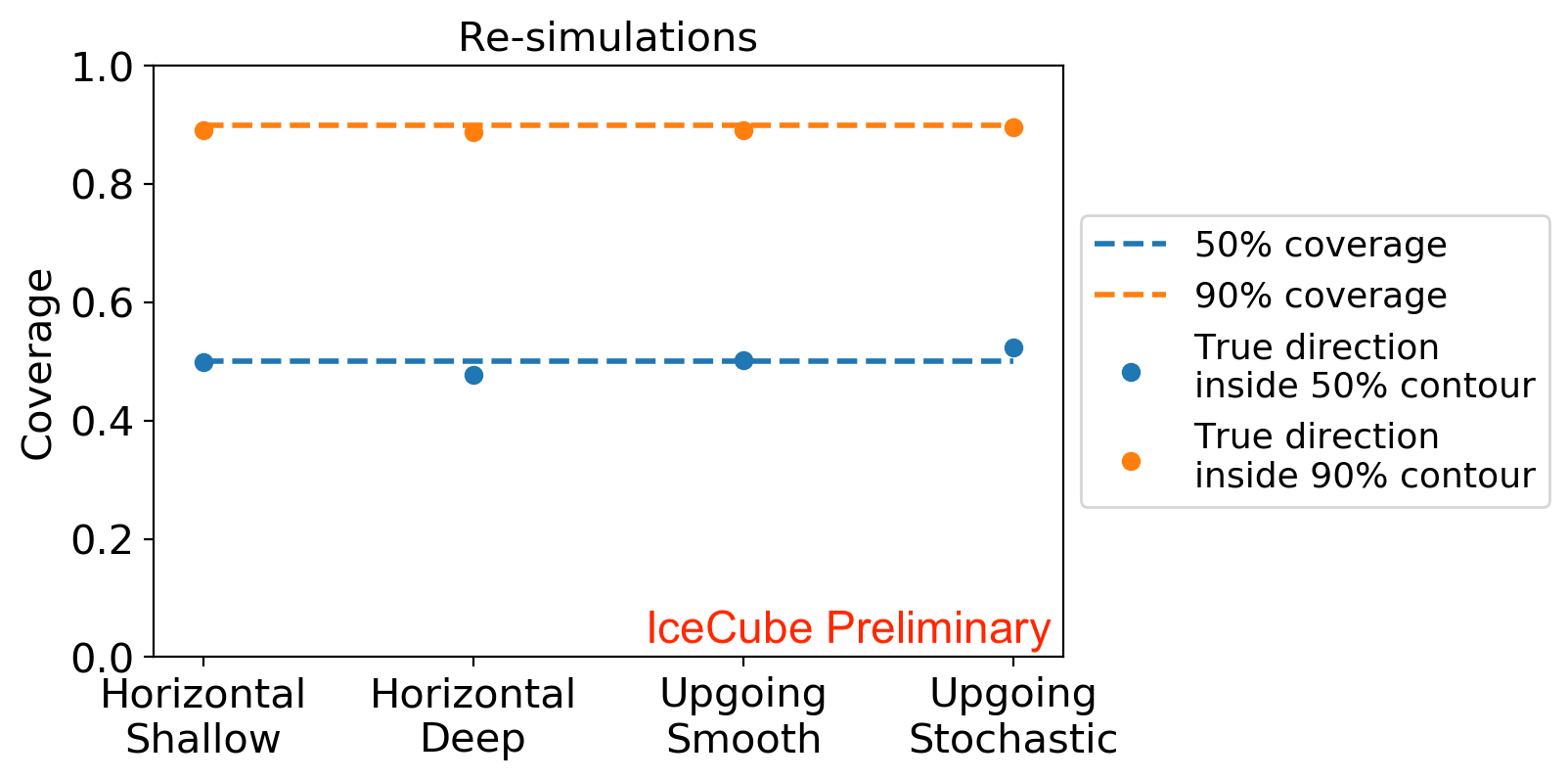}
  \caption{}
  \label{fig:coverage1}
\end{subfigure}%
\begin{subfigure}{.5\textwidth}
  \centering
  \includegraphics[width=\linewidth]{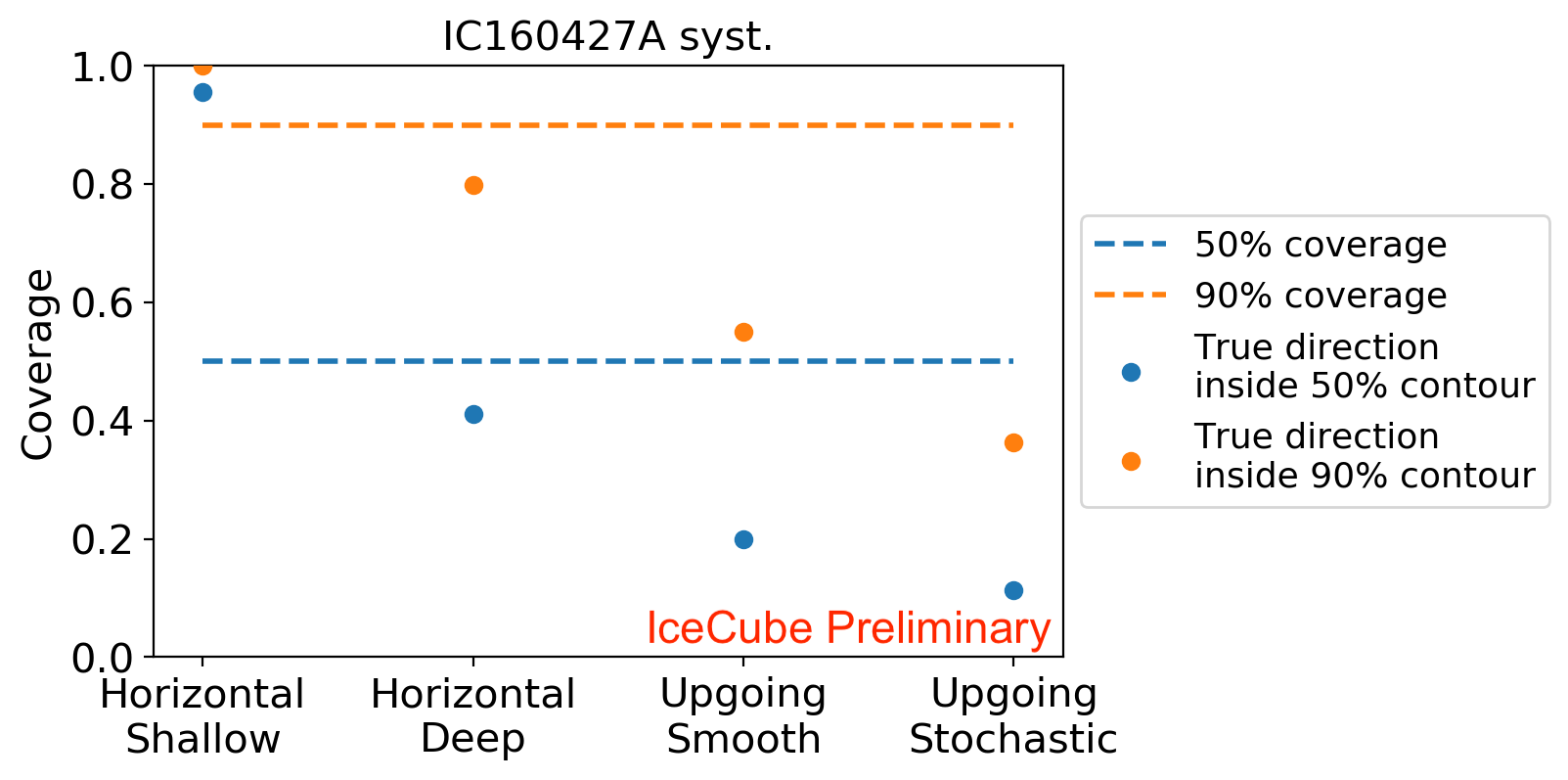}
  \caption{}
  \label{fig:coverage2}
\end{subfigure}
\caption{Study of the coverage of the contours calculated with the re-simulations of each category (\ref{fig:coverage1}) and with the IC160427A correction values (\ref{fig:coverage2}). The dashed lines indicate that for 50\% (90\%) of the reconstructed events the simulated direction lies within the 50\% (90\%) error contour. The dots represent the actual percentage of events that satisfy this condition.}
\label{fig:coverage}
\end{figure}

To measure the performance of this method, we study the coverage of the contours by dividing the events within each category into two sets: one to calculate the correction values and the other to evaluate them. The evaluation set is used to obtain the amount of events whose simulated direction is inside the error contours calculated with the values from the calculation set. This is repeated by randomly dividing the re-simulations to find the average 50\% and 90\% coverage (Figure~\ref{fig:coverage1}). To compare the performance of the original correction values, the coverage is evaluated again using the contours calculated with the IC160427A re-simulations (Figure~\ref{fig:coverage2}). In both cases, accurate contours correspond to the localisation of blue (orange) dots around the blue (orange) dashed line.  

\begin{figure}[h] 
  \begin{subfigure}[b]{0.5\linewidth}
    \centering
    \includegraphics[width=\linewidth]{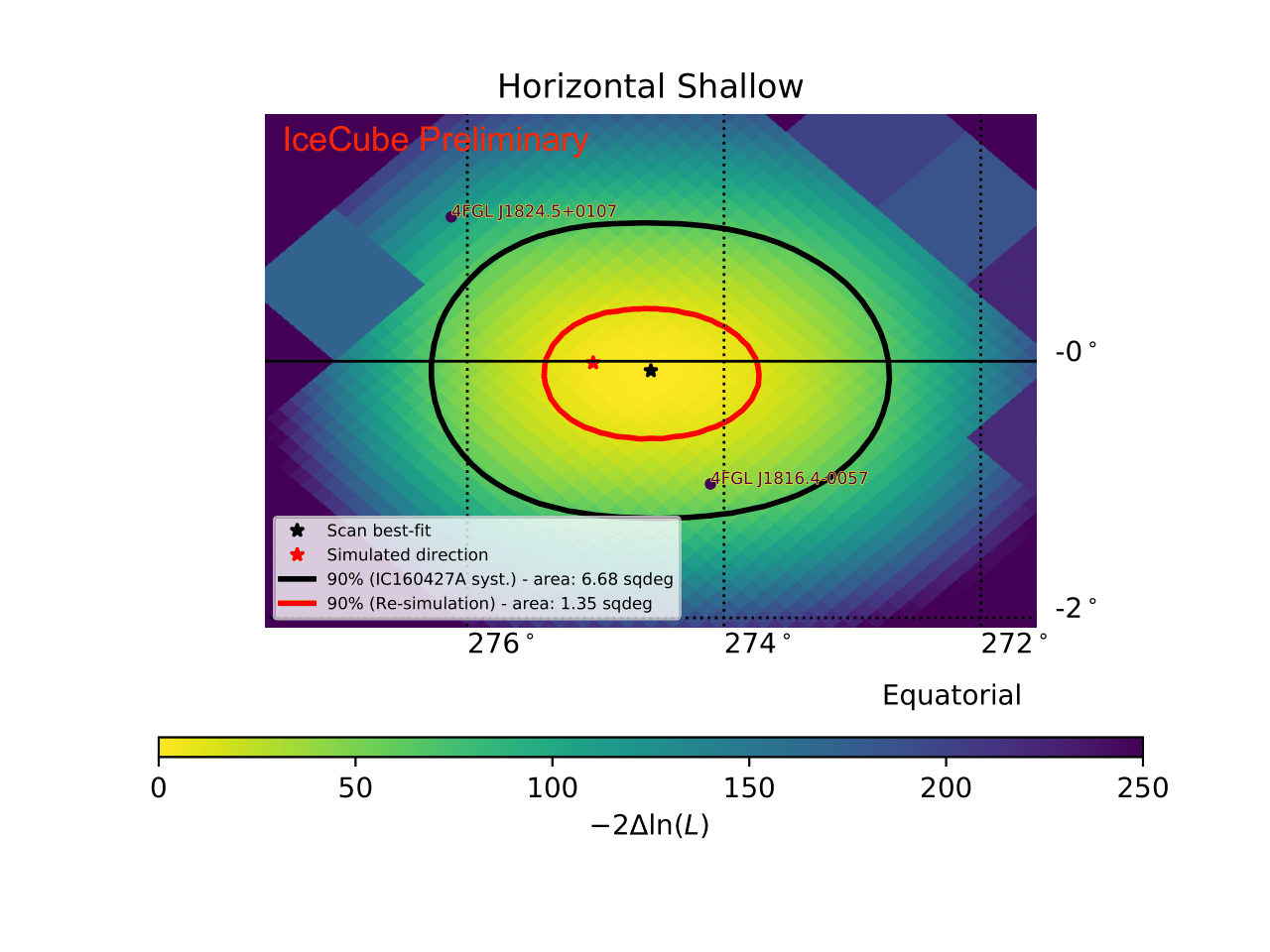} 
    \label{fig7:a} 
    \vspace{-7ex}
  \end{subfigure}
  \begin{subfigure}[b]{0.5\linewidth}
    \centering
    \includegraphics[width=\linewidth]{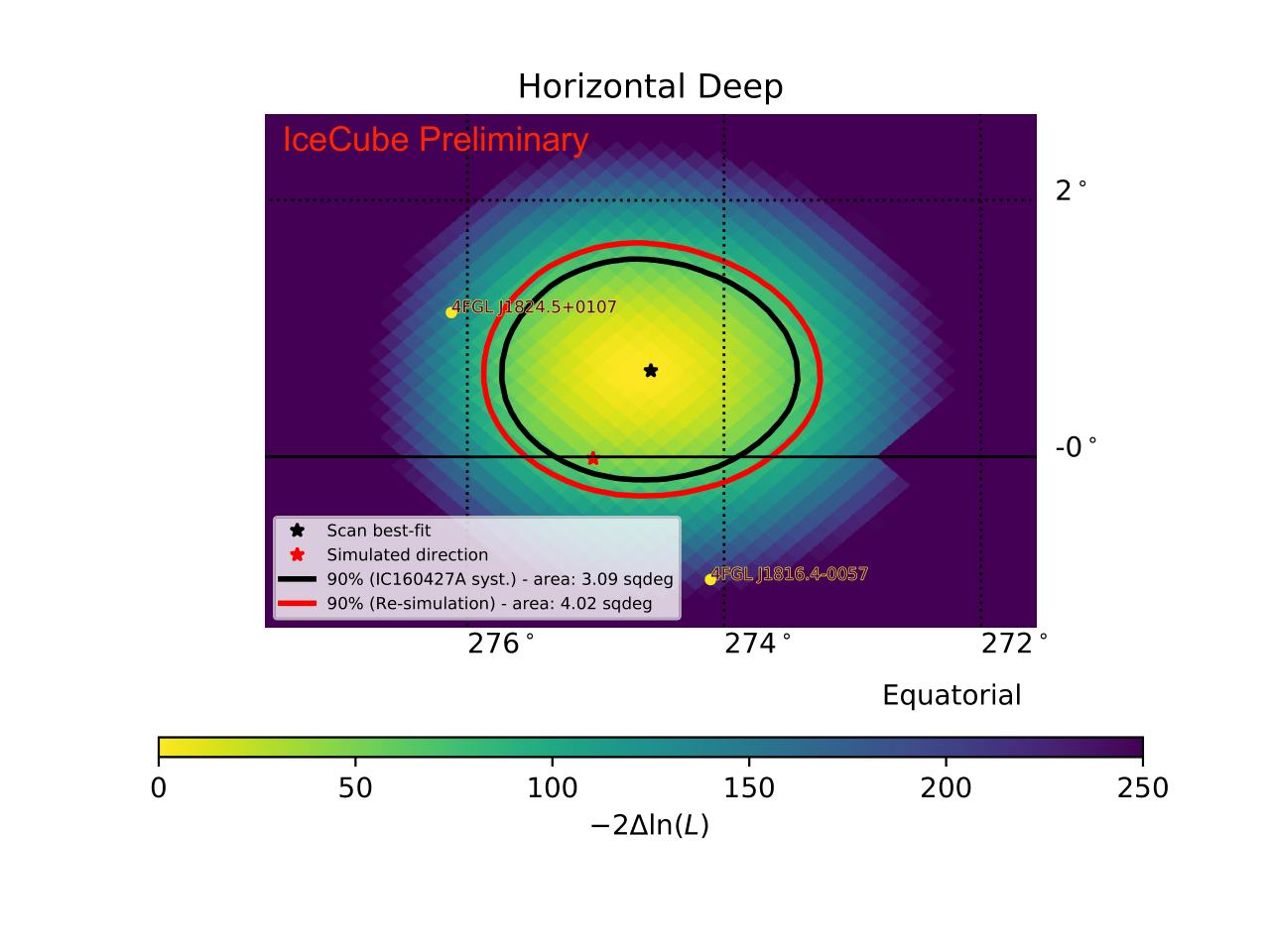} 
    \label{fig7:b} 
    \vspace{-7ex}
  \end{subfigure} 
  \begin{subfigure}[b]{0.5\linewidth}
    \centering
    \includegraphics[width=\linewidth]{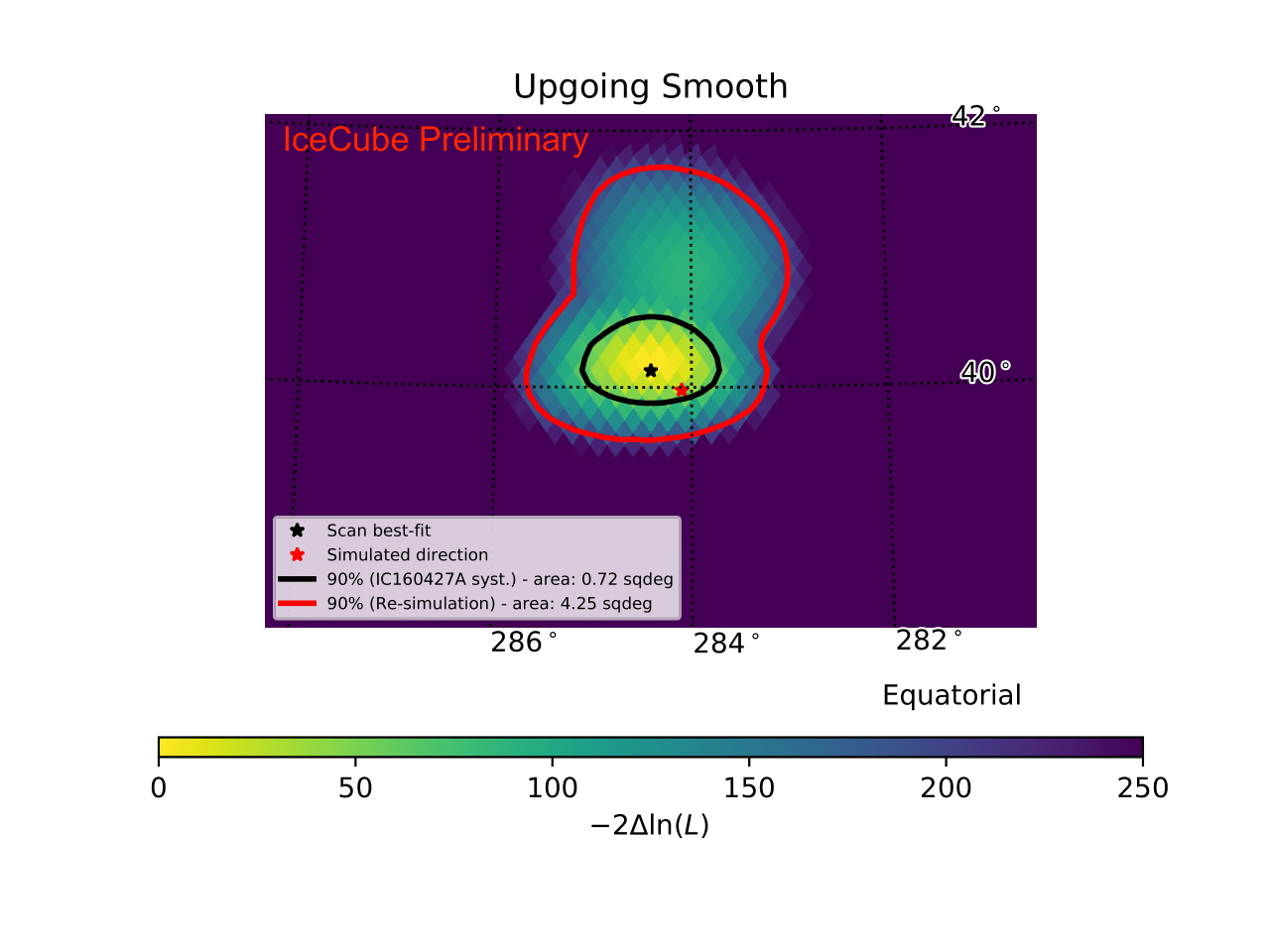} 
    \label{fig7:c} 
    \vspace{-7ex}
  \end{subfigure}
  \begin{subfigure}[b]{0.5\linewidth}
    \centering
    \includegraphics[width=\linewidth]{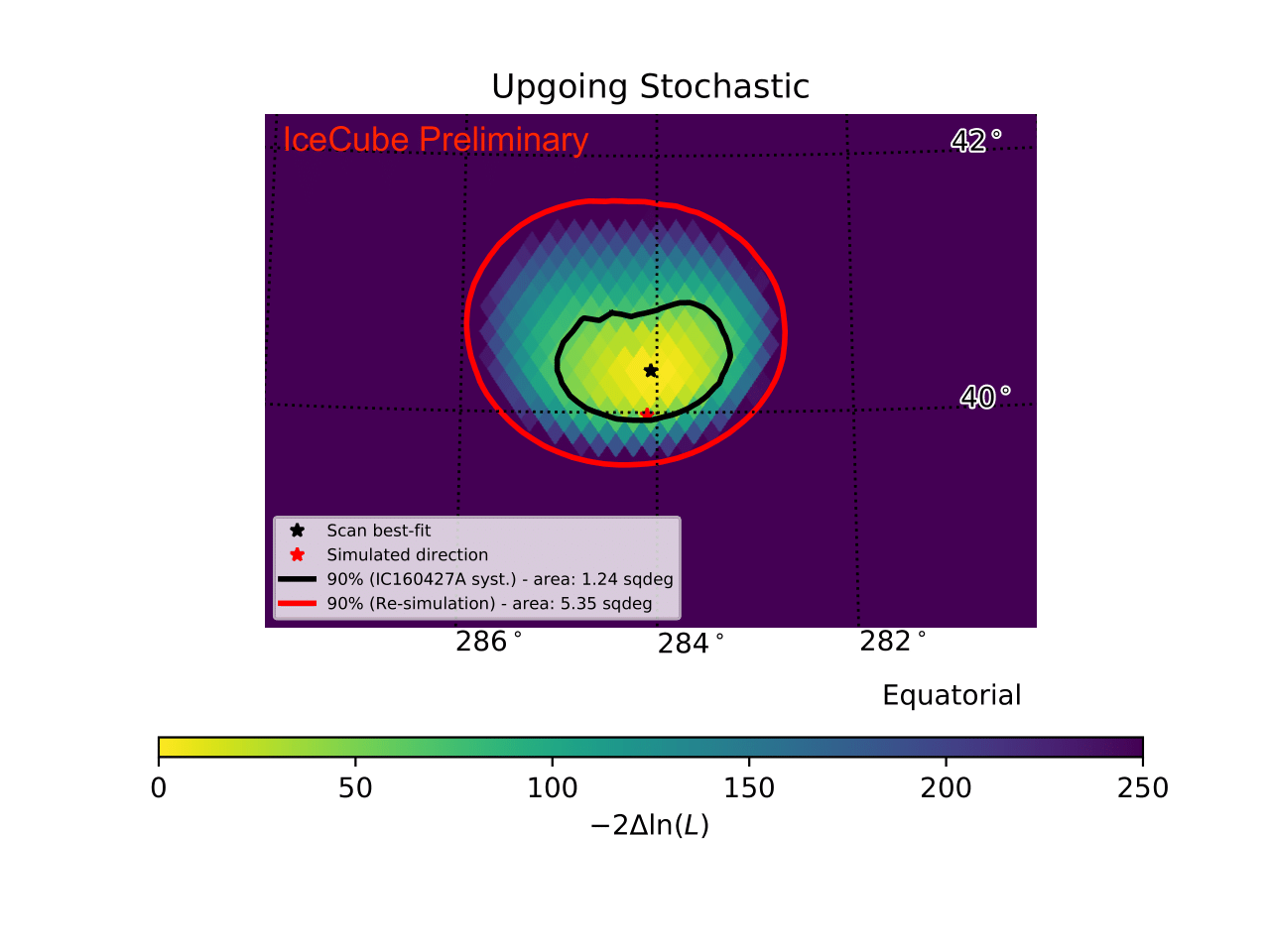} 
    \label{fig7:d} 
    \vspace{-7ex}
  \end{subfigure} 
  \caption{Results of the \textit{Millipede} scan for one random event of each category. The black star represents the scan best fit and the red star is the true direction. The black line is the 90\% contour calculated with the IC160427A correction values (currently used for every IceCube's real-time alert) and the red line is the 90\% contour calculated with the correction values of the event's category (see Table~\ref{tab:status}).}
  \label{fig:contours} 
\end{figure}

Lastly, the error contours calculated using IC160427A (black) and the new re-simulations (red) for a random event of each category are shown in Figure~\ref{fig:contours}. This expresses that the area of the contour ultimately depends on the steepness and shape of the likelihood landscape, so the scaling will have a different impact for each alert. It also means that, while we now know that using the same parameters is not correct, on average it is a good approximation to our new, more reliable error contours. It must be noted that it is not possible to calculate "true" error contours that would account for all existing systematic uncertainties because some of them are known but not implemented in the simulation software whereas some are completely unknown. 

\section{Conclusion and outlook}\label{sec:outlook}

This on-going analysis has studied the validity of the correction values currently used in IceCube's real-time program to derive the angular uncertainty of alerts, arriving to the conclusion that a more robust treatment of the systematic uncertainties is needed. The first step to solving this problem was to create simple categories of muons that represent most real-time neutrino alerts. 100 muons of each class were re-simulated varying the ice model parameters with a novel simulation tool. The re-simulations were then reconstructed using a complex and robust reconstruction method called \textit{Millipede}. The resulting scans containing likelihood landscapes were used to obtain new correction values that are used to derive error contours with the expected coverage in future real-time alerts. 

Moving forward with this analysis, more re-simulations will be prepared. The current categories will continue to be scanned in order to reach enough statistics ($\mathcal{O}(100)$ re-simulations) and new classes will be defined to cover all possible tracks, including but not limited to various zenith angles, energy losses, initial neutrino energies and depths of the track in the detector. The aim is to apply specific correction values to each new neutrino alert by interpolating between the different categories. 

Lastly, the re-simulations will be reconstructed using other reconstruction methods (like those described in~\cite{abbasi2021muontrack}) that are faster, simpler and less computationally expensive. This will enable us to compare their performance on high-energy neutrinos and eventually decide whether \textit{Millipede} should continue to be used in the real-time program.

\bibliographystyle{ICRC}
\bibliography{skeleton}

\clearpage
\section*{Full Author List: IceCube Collaboration}




\scriptsize
\noindent
R. Abbasi$^{17}$,
M. Ackermann$^{59}$,
J. Adams$^{18}$,
J. A. Aguilar$^{12}$,
M. Ahlers$^{22}$,
M. Ahrens$^{50}$,
C. Alispach$^{28}$,
A. A. Alves Jr.$^{31}$,
N. M. Amin$^{42}$,
R. An$^{14}$
% ...
\\

\noindent
% ...
$^{12}$ Universit{\'e} Libre de Bruxelles, Science Faculty CP230, B-1050 Brussels, Belgium \\
% ...
$^{14}$ Department of Physics and Laboratory for Particle Physics and Cosmology, Harvard University, Cambridge, MA 02138, USA \\
% ...
$^{17}$ Department of Physics, Loyola University Chicago, Chicago, IL 60660, USA \\
$^{18}$ Dept. of Physics and Astronomy, University of Canterbury, Private Bag 4800, Christchurch, New Zealand \\
% ...
$^{22}$ Niels Bohr Institute, University of Copenhagen, DK-2100 Copenhagen, Denmark \\
% ...
$^{28}$ D{\'e}partement de physique nucl{\'e}aire et corpusculaire, Universit{\'e} de Gen{\`e}ve, CH-1211 Gen{\`e}ve, Switzerland \\
% ...
$^{31}$ Karlsruhe Institute of Technology, Institute for Astroparticle Physics, D-76021 Karlsruhe, Germany  \\
% ...
$^{42}$ Bartol Research Institute and Dept. of Physics and Astronomy, University of Delaware, Newark, DE 19716, USA \\
% ...
$^{50}$ Oskar Klein Centre and Dept. of Physics, Stockholm University, SE-10691 Stockholm, Sweden \\
% ...
$^{59}$ DESY, D-15738 Zeuthen, Germany \\
% ...

\subsection*{Acknowledgements}

\noindent
USA {\textendash} U.S. National Science Foundation-Office of Polar Programs,
U.S. National Science Foundation-Physics Division,
U.S. National Science Foundation-EPSCoR,
Wisconsin Alumni Research Foundation,
Center for High Throughput Computing (CHTC) at the University of Wisconsin{\textendash}Madison,
Open Science Grid (OSG),
Extreme Science and Engineering Discovery Environment (XSEDE),
Frontera computing project at the Texas Advanced Computing Center,
U.S. Department of Energy-National Energy Research Scientific Computing Center,
Particle astrophysics research computing center at the University of Maryland,
Institute for Cyber-Enabled Research at Michigan State University,
and Astroparticle physics computational facility at Marquette University;
Belgium {\textendash} Funds for Scientific Research (FRS-FNRS and FWO),
FWO Odysseus and Big Science programmes,
and Belgian Federal Science Policy Office (Belspo);
Germany {\textendash} Bundesministerium f{\"u}r Bildung und Forschung (BMBF),
Deutsche Forschungsgemeinschaft (DFG),
Helmholtz Alliance for Astroparticle Physics (HAP),
Initiative and Networking Fund of the Helmholtz Association,
Deutsches Elektronen Synchrotron (DESY),
and High Performance Computing cluster of the RWTH Aachen;
Sweden {\textendash} Swedish Research Council,
Swedish Polar Research Secretariat,
Swedish National Infrastructure for Computing (SNIC),
and Knut and Alice Wallenberg Foundation;
Australia {\textendash} Australian Research Council;
Canada {\textendash} Natural Sciences and Engineering Research Council of Canada,
Calcul Qu{\'e}bec, Compute Ontario, Canada Foundation for Innovation, WestGrid, and Compute Canada;
Denmark {\textendash} Villum Fonden and Carlsberg Foundation;
New Zealand {\textendash} Marsden Fund;
Japan {\textendash} Japan Society for Promotion of Science (JSPS)
and Institute for Global Prominent Research (IGPR) of Chiba University;
Korea {\textendash} National Research Foundation of Korea (NRF);
Switzerland {\textendash} Swiss National Science Foundation (SNSF);
United Kingdom {\textendash} Department of Physics, University of Oxford.

\end{document}